%% file: acl_latex.tex
\pdfoutput=1

\RequirePackage{snapshot}
\documentclass[11pt]{article}

\input{preamble}

\usepackage[preprint]{acl}

\usepackage{times}
\usepackage{latexsym}

\usepackage[T1]{fontenc}

\usepackage[utf8]{inputenc}

\usepackage{microtype}

\usepackage{inconsolata}

\usepackage{graphicx}

\title{\titletext}

\author{Dhananjay Ashok\textsuperscript{1}\thanks{Work done during an internship at Amazon.} \quad
  Suraj Nair\textsuperscript{2} \quad Mutasem Al-Darabsah\textsuperscript{2} \\ 
  \textbf{Choon Hui Teo\textsuperscript{2} \quad  Tarun Agarwal\textsuperscript{2} \quad  Jonathan May\textsuperscript{2}} \\ 
  \textsuperscript{1}Information Sciences Institute, University of Southern California \quad \textsuperscript{2}Amazon \\
  \texttt{ashokd@isi.edu} \quad \texttt{\{srjnair, mutasema, choonhui, tagar, jnatmay\}@amazon.com}\\}

\begin{document}
\maketitle
\begin{abstract}
\input{sections/0_abstract}
\end{abstract}

\input{content}

\bibliography{main}

\appendix
\input{sections/A_appendix}

\end{document}

%% file: preamble.tex
\usepackage{graphicx}
\usepackage{booktabs}
\usepackage{multirow}
\usepackage{xspace}

\newcommand{\titletext}{A Representation Sharpening Framework for Zero Shot Dense Retrieval}

\newcommand{\method}{representation sharpening}
\usepackage{amsmath,amsfonts,bm}

\usepackage{xparse}

\newcommand{\corpus}{D}
\newcommand{\references}{D'}
\newcommand{\docindex}{\mathbf{D}}
\NewDocumentCommand{\doc}{g}{
  d
  \IfValueT{#1}{_{#1}}%
  \xspace
}
\NewDocumentCommand{\otherdoc}{g}{
  \doc'
  \IfValueT{#1}{_{#1}}%
  \xspace
}
\NewDocumentCommand{\query}{g}{
  q
  \IfValueT{#1}{_{#1}}%
  \xspace
}
\NewDocumentCommand{\queryset}{g}{
  Q
  \IfValueT{#1}{_{#1}}%
  \xspace
}
\NewDocumentCommand{\queryembedset}{g}{
  \mathbf{Q}
  \IfValueT{#1}{_{#1}}%
  \xspace
}
\NewDocumentCommand{\docembed}{g}{
  \mathbf{d}
  \IfValueT{#1}{_{#1}}%
  \xspace
}
\NewDocumentCommand{\queryembed}{g}{
  \mathbf{q}
  \IfValueT{#1}{_{#1}}%
  \xspace
}
\newcommand{\simscore}{\text{s}\xspace}

\def\eqref#1{equation~\ref{#1}}

\def\1{\bm{1}}

\DeclareMathAlphabet{\mathsfit}{\encodingdefault}{\sfdefault}{m}{sl}
\SetMathAlphabet{\mathsfit}{bold}{\encodingdefault}{\sfdefault}{bx}{n}



%% file: sections/0_abstract.tex
Zero-shot dense retrieval is a challenging setting where a document corpus is provided without relevant queries, necessitating a reliance on pretrained dense retrievers (DRs). However, since these DRs are not trained on the target corpus, they struggle to represent semantic differences between similar documents. 
To address this failing, we introduce a training-free \textbf{\method} framework that augments a document's representation with information that helps differentiate it from similar documents in the corpus. 
On over twenty datasets spanning multiple languages, the \method{} framework proves consistently superior to traditional retrieval, setting a new state-of-the-art on the BRIGHT benchmark. 
We show that \method{} is compatible with prior approaches to zero-shot dense retrieval and consistently improves their performance.
Finally, we address the performance-cost tradeoff presented by our framework and devise an indexing-time approximation that preserves the majority of our performance gains over traditional retrieval, yet suffers no additional inference-time cost. 

%% file: content.tex
\input{sections/1_introduction}

\input{sections/related_work}

\input{sections/2_method}

\input{sections/3_0_experiments}

\input{sections/5_approximation}

\input{sections/4_ablations}

\input{sections/conclusion}

\input{sections/limitations}

%% file: sections/1_introduction.tex
\section{Introduction}
\label{sec:introduction}
Dense retrieval systems represent queries and documents in a semantic space, using the similarity of their embeddings as an estimate of a document's relevance to a query~\citep{yih-etal-2011-learning}. This approach has been shown to work well in data-rich domains like Question Answering~\citep{karpukhin-etal-2020-dense} and Web Search~\citep{mao-etal-2022-convtrans}. However, in more realistic scenarios, we do not have access to relevant queries for the documents in our corpus~\citep{dai2023promptagator} and must perform zero-shot retrieval~\citep{gao-etal-2023-precise}. 
Approaches to this setting rely on generic, pretrained DRs, making alignment to the target corpus a major concern~\citep{thakur2021beir,wang2023querydoc}. Such alignment often requires creating synthetic datasets for finetuning~\citep{ma-etal-2021-zero}, incentivizing the DR to create document embeddings that are similar to those of relevant queries. While often successful~\citep{izacard2022unsupervised}, retraining risks catastrophic forgetting~\citep{goodfellow2013empirical}, can be computationally infeasible~\citep{zhang2025qwen3}, and, in the case of closed-source DRs~\citep{neelakantan2022text, anthropicEmbeddingsAnthropic} is impossible. Training-free adaptation may avoid these limitations. 
\input{figures/coninf}

We introduce a framework of \textbf{\method} (Figure~\ref{fig:confing}), that augments a document's embedding during inference to enhance a DR without retraining. To achieve this, we first generate synthetic queries either using standard query generation~\citep{ma-etal-2021-zero} or, with our novel \textbf{contrastive query generation} paradigm, which creates queries that contain information on what distinguishes a document from other, similar documents in the target corpus. During inference, we use the test query to assign weights to each generated query, and mix them into the document embeddings before computing the relevance score. This accentuates the unique aspects of each document, sharpening its representation with respect to the test query and enabling more precise retrieval. 

On six datasets from the BEIR benchmark~\citep{thakur2021beir}, \method{} outperforms traditional inference and a document expansion baseline, improving NDCG@10 by 6.9\% on average. 
We apply our framework to prior approaches to training-free, zero-shot dense retrieval, and show that it consistently improves all existing methods. Beyond standard benchmarks, we apply our framework to the complex reasoning-based BRIGHT~\cite{su2025bright} benchmark, boosting pre-trained DRs and achieving a new state-of-the-art on eight of eleven measured subsets. We further show that our method is applicable to multiple languages, and improves the performance of DRs on the Hindi, Swahili, Korean and Thai splits of the MIRACL benchmark \cite{zhang-etal-2023-miracl}. 

Finally, we address implementation costs and present an alternative approach that, instead of adapting the inference procedure, performs a one-time alteration of the vector embeddings of the documents in the index. This indexing-time approach proves consistently superior to traditional retrieval and a document expansion baseline, showing that contrastive queries can boost retrieval while incurring \textbf{no additional inference-time cost}. 

%% file: figures/coninf.tex
\begin{figure*}[t]
    \centering
    \includegraphics[width=\linewidth]{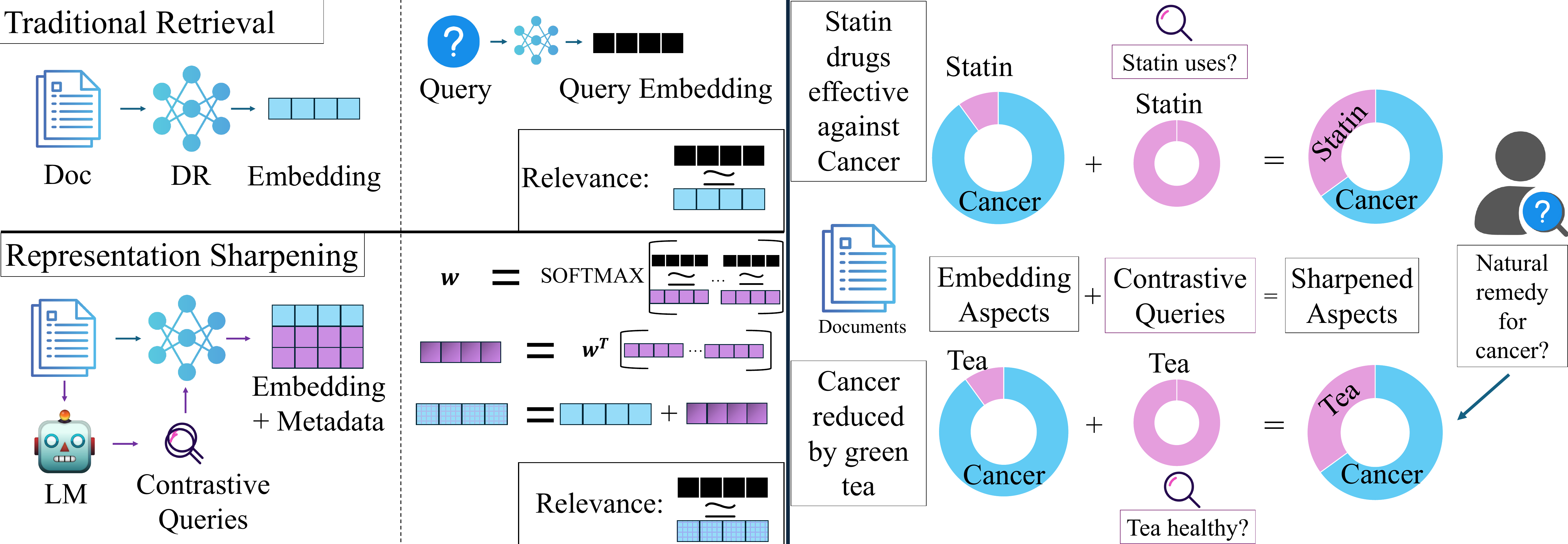}
    \caption{\textbf{Left}: Comparison of the traditional dense retrieval pipeline and our proposed framework for \textbf{\method}. During the indexing phase, we generate contrastive queries and store them along with a document's embedding as metadata. During inference, the inference query is used to determine weights that signify which contrastive queries are most relevant and we aggregate their embeddings using these weights. We use this aggregated embedding to augment the document representation before computing relevance. \textbf{Right}: Visualization of how sharpening improves retrieval. The contrastive queries help refine the embeddings and accentuate the most unique aspects of each document, hence enabling more precise differentiation between similar documents.}
    \label{fig:confing}
\end{figure*}

%% file: sections/related_work.tex
\section{Background and Related Work}
\label{sec:related-work}
Statistical approaches to document indexing and retrieval have a long history~\citep{bookstein1974probabilistic, ponte1998language}, with most work assuming access to plentiful training data in the form of document-query relevance annotations~\citep{ bajaj2016ms, kwiatkowski-etal-2019-natural}. However, while it has long been acknowledged that such a setting is limiting~\citep{croft1979using}, performing retrieval zero-shot, i.e., without relevance labels was a historically challenging problem~\citep{Harman1992RelevanceFR, thakur2021beir}. Modern methods tuned DRs on large training corpora and performed Zero-Shot transfer between domains~\citep{yu-etal-2022-coco, wang-etal-2022-gpl, lin-etal-2023-maggretriever}. This approach relies on the target domain being well aligned with the training set~\citep{chandradevan-etal-2024-duqgen}, which, as pointed out by ~\citet{izacard2022unsupervised} and ~\citet{gao-etal-2023-precise}, can rarely be assumed, sparking an interest in alternate approaches~\citep{dai2023promptagator}.  

The emergence of powerful LMs~\citep{raffel2020exploring} with impressive few-shot~\citep{brown2020language} and instruction-following~\citep{ouyang2022training} capabilities presented such an alternative. Significant gains have been made by leveraging these LMs to generate synthetic queries~\citep{ma-etal-2021-zero, dai2023promptagator} and documents~\citep{ma2023pre,li2024syneg} that can be used to train target-domain specific DRs. However, not only is adjusting the weights of DRs an often computationally restrictive process~\citep{zhang2025qwen3}, it is also prone to removing useful, pre-existing knowledge~\citep{goodfellow2013empirical}. Additionally, for increasingly superior closed-source embedding systems~\citep{neelakantan2022text, anthropicEmbeddingsAnthropic}, weights are inaccessible, making fine-tuning impossible. 

These concerns have prompted the development of training-free approaches~\citep{gao-etal-2023-precise} to enhancing pretrained DRs on unseen test domains. Recent methods~\citep{wang2023querydoc, shen2023large} enhance the capabilities of DRs by expanding the inference queries to include more explicit information. However, this requires continuous access to an LM during the inference phase, a prohibitively expensive design decision~\citep{MLSYS2022_462211f6}. In this work, we take an alternate approach and propose a solution that only requires LM access during the indexing phase of retrieval, increasing performance without relevance labels, retraining or significantly increasing deployment costs.  

\noindent\textbf{Query Generation:}
Synthetic query generation has proven a powerful approach to creating training sets for DRs~\citep{ma-etal-2021-zero, sachan-etal-2022-improving, bonifacio2022inpars}. Subsequent work has focused on scaling the technique~\citep{wang-etal-2024-improving-text}, improving filtering~\citep{almeida-matos-2024-exploring}, incorporating domain knowledge~\citep{xia2025knowledge} and capturing a range of possible user intents~\citep{lee-etal-2024-disentangling}. Despite this progress, there has been little change to the fundamentally one-to-many paradigm of query generation, i.e., prior work generates queries that are relevant (or not relevant~\citep{lv2015negative}) to a single document in the corpus. In this work, we question this practice, proposing instead a \textbf{many-to-many} paradigm, that generates contrastive queries which are relevant to some documents, and simultaneously not relevant to other, similar documents in the corpus. This paradigm is orthogonal to specific improvements in query generation practices (e.g., incorporating additional information~\citep{gupta-etal-2025-schema}), as these methods can easily be applied within our paradigm as well. To the best of our knowledge, the only other works to explore such a paradigm do so for evaluation~\citep{weller2024nevir} and cross-language training dataset construction~\citep{mayfield2023synthetic}, while we are the first to show that such a framework for query generation can be used to actively improve performance without retraining.

We summarize our contributions as: 
\begin{enumerate}
    \item A zero-shot framework for \textbf{\method}, that boosts pretrained DRs on a wide range of datasets and languages. 
    \item \textbf{Contrastive query generation}, which creates queries that accentuate the aspects that most distinguish a document from other, similar documents within corpus. 
    \item An indexing-time method that directly alters the document embeddings in the index, and boosts retrieval performance while incurring \textbf{no additional inference-time cost}. 
\end{enumerate}

%% file: sections/2_method.tex
\input{tables/beir}
\section{Representation Sharpening Framework}
\label{sec:method}

In this section, we first introduce the traditional paradigm of inference for dense retrieval and then provide a high-level overview of our alternative framework for \textbf{\method}. Finally, we explain how we operationalize this framework using LMs for contrastive query generation.

\subsection{Traditional Retrieval}
Consider a text corpus of documents  $\corpus = \{\doc{1}, \doc{2}\ldots\}$ that must be ranked for relevance with respect to an inference query with text $q$. The traditional retrieval pipeline starts by pre-computing the index of document representations $\docindex = \{\docembed{1}, \docembed{2}, \ldots \}$, where $\docembed{i} \in\mathbb{R}^m$ is the semantic representation of $\doc{i}$ under some pretrained DR with embedding size $m$. During inference, we compute the representation $\queryembed\in\mathbb{R}^m$ of the query and rank the documents based on $\simscore: \mathbb{R}^m\times\mathbb{R}^m\to\mathbb{R}$, a measure of the similarity between the query and document representations. For instance, when using the standard measure of cosine similarity, the score of document $d$ is computed as $\simscore({\mathbf{q}}, \mathbf{d}) = \frac{\mathbf{q} \cdot \mathbf{d}}{\|\mathbf{q}\|\|\mathbf{d}\|}$

When two documents have similar representations $\docembed{1}, \docembed{2}$, it is usually the case that if query representation $\queryembed$ is similar to $\docembed{1}$, it will also be similar to $\docembed{2}$. In such cases, if traditional retrieval inference is to determine which of the documents is \textbf{more} relevant, the representations must properly encode the subtle semantic differences between the two documents. However, this is often not the case for pretrained DRs~\citep{ren-etal-2023-thorough}. 

\subsection{Representation Sharpening}
We propose that instead of exclusively relying on the DR to represent these differences, we explicitly provide this information in the form of \textbf{contrastive queries}. Given a document $\doc$, the set of contrastive queries $\queryset{\doc}$ are the queries for which $d$ is relevant, and some other, similar documents in the corpus are \textbf{not} relevant. We compute their representations $\queryembedset{\doc}=\{\queryembed{d, 1}, \queryembed{d, 2} \ldots\}$ and store them as meta-data for document $\doc$ in the index. During inference, we shift the document embedding in the direction of the contrastive query embeddings with:
\begin{equation}
    \docembed^* = \docembed + \alpha \cdot g(\queryembed, \queryembedset{\doc})
    \label{eq:main}
\end{equation}

\noindent where $g$ is an inference-query aware aggregation function over the contrastive query embeddings. In our work, we consider $g$ to be a convex combination of the contrastive query embeddings, where the weights are determined by how similar they are to the inference query, as judged by a softmax: 
\begin{equation}
    g(\queryembed, \queryembedset{d}) = \sum_{\queryembed{d, i}\in\queryembedset{d}}\frac{\exp(\simscore(\queryembed, \queryembed{d, i}))}{\sum_{\queryembed{d, j}\in\queryembedset{d}}\text{exp}(\simscore(\queryembed, \queryembed{d, j}))}\queryembed{d, i}
    \label{eq:softmax}
\end{equation}

This `sharpens'~\citep{berthelotmixmatch2019, huang2025selfimprovement} the document's representation, and reinforces the aspects that most saliently distinguish it from the other documents in the corpus in a way that is relevant to the inference query at hand. The relevance score of $\doc$ is then $\simscore(\queryembed, \docembed^*)$.  

\subsection{Selecting Contrastive References}
To facilitate the framework described above, given a document $\doc$, we must generate contrastive queries that differentiate between $\doc$ and other, similar documents in the corpus. To do so, we must select the similar documents that will serve as \textit{contrastive references} for $\doc$. A natural choice is to leverage the DR and select the nearest neighbors of $\doc$ in the document index $\docindex$. However, a document may simultaneously address a variety of topics, and selecting the nearest neighbors risks a failure to capture the full diversity of a document's topic range. Instead, we subsample a large, local neighborhood of $\doc$ in $\docindex$ and identify groups using unsupervised clustering algorithms~\citep{aggarwal1999merits}, e.g., KMeans~\citep{Jin2010}. We then select one document $\otherdoc$ from each cluster to form our set of contrastive references $\references_{\doc} = (\otherdoc{1}, \otherdoc{2}, \ldots)$.

\subsection{Generating Contrastive Queries}
\label{sec:generating}
For a given document $\doc$ and set of contrastive references $\references_{\doc}$,  we instruct an LM to create queries that are relevant to $d$, but irrelevant to some $(\otherdoc{i}, \otherdoc{j}, \ldots)\subseteq\references_{\doc}$. This is a departure from existing approaches to query generation~\citep{ma-etal-2021-zero, sachan-etal-2022-improving}, that perform one-to-many generation, i.e., generating queries for one document at a time. Under the standard paradigm, it is unclear whether a relevant query generated for $\doc$ is also relevant to $\otherdoc$, and can produce generic queries that do not accentuate the unique contents of $\doc$. For instance, given either document in Figure~\ref{fig:confing} (right), standard query generation may produce `How to prevent cancer?'. However, this is generic, as it is applicable to both documents. We instead propose a \textbf{many-to-many} paradigm for query generation that allows us to create more precise queries that highlight unique document aspects. As shown in Section~\ref{sec:experiments}, these precise queries prove more capable of separating a document from other, similar documents in the corpus.

%% file: tables/beir.tex
\begin{table*}[th]
\centering
\begin{tabular}{@{}llrrrrrrr@{}}\toprule
DR         & Variant         & FiQA           & NFCorpus       & SciFact        & T-COV     & SciDocs        & Arguana        & Avg            \\ \midrule
Contriever & Trad            & 24.19          & 31.23          & 57.05          & 23.54          & 14.38          & 49.43          & 33.30           \\
           & Doc2Query       & 20.14          & 31.71          & 57.20           & 28.47          & 14.40          & 50.29         & 33.70          \\
           & SimSharp (Ours)      & 29.45          & 32.81          & 65.61          & 33.25          & 17.24          & 53.75          & 38.69          \\
           & ConSharp (Ours) & \textbf{30.21} & \textbf{34.89} & \textbf{68.51} & \textbf{35.36} & \textbf{17.94} & \textbf{57.06} & \textbf{40.67} \\\midrule
Qwen3      & Trad            & 34.28          & 30.41          & 64.61          & 58.49          & 16.48          & 66.60           & 45.15          \\
           & Doc2Query       & 35.29          & 30.83          & 64.65          & \textbf{68.52}          & 20.06          & 65.02          & 47.40           \\
           & SimSharp (Ours)      & 38.90           & 33.52          & 70.51          & 65.26 & 21.78          & 66.18          & 49.35          \\
           & ConSharp (Ours) & \textbf{40.62} & \textbf{34.30}  & \textbf{71.88} & 62.90           & \textbf{22.55} & \textbf{69.97}          & \textbf{50.37} \\\midrule
E5-Mistral & Trad            & 36.89          & 25.75          & 64.79          & 44.66          & 2.69           & 50.16          & 37.49          \\
           & Doc2Query       & 34.62          & 20.07          & 64.89          & 39.81          & 6.36           & 47.59          & 35.55          \\
           & SimSharp (Ours)      & 43.08          & 33.78          & 72.23          & \textbf{50.73} & 7.07           & 57.03          & 43.98          \\
           & ConSharp (Ours) & \textbf{45.15} & \textbf{35.51} & \textbf{74.33} & 50.25          & \textbf{8.57}  & \textbf{59.42}          & \textbf{45.54}          \\ \bottomrule
\end{tabular}
\caption{NDCG@10 on BEIR benchmark when comparing traditional retrieval (Trad) to our proposed representation sharpening approach (SimSharp and ConSharp). Across datasets and underlying dense retrievers, ConSharp outperforms both traditional inference and a Doc2Query baseline with an average improvement of 6.9\%. }
\label{table:beir}
\end{table*}

%% file: sections/3_0_experiments.tex
\input{sections/3_1_beir}

\input{sections/3_2_bright}

\input{sections/3_3_miracl}

%% file: sections/3_1_beir.tex
\section{Outperforming Traditional Inference}
\label{sec:experiments}
We use six datasets from the BEIR benchmark, spanning the domains of financial QA (FiQA), scientific understanding (SciDocs, SciFACT, NFCorpus), scientific news (Trec-COVID) and counter argument mining (Arguana). We select some of the most widely used pretrained DRs: Contriever~\cite{izacard2022unsupervised}, Qwen3Embedding-0.6B~\citep{zhang2025qwen3} and E5-Mistral~\cite{wang2022text} and repeat the following experiment: 

\input{tables/prior}

\noindent\textbf{Contrastive Reference Selection:} Given a corpus $\corpus$ and pretrained DR, we compute the document index $\docindex$. For each document in the index, we subsample the top 100 nearest neighbors (by cosine similarity) and then cluster them using KMeans. Following standard practices~\citep{shahpure2020clustering}, we select the number of clusters $k^*\in[3, 10]$ which achieves the highest silhouette score $ \sum_{i=1}^{100}\frac{b_i - a_i}{\max(a_i, b_i)}$, where $a_i$ is the average distance of neighbor $i$ to other points in the same cluster and $b_i$ is the minimum average distance of neighbor $i$ to other clusters. This allows us to select a variable number of clusters for each document based on the range of topics it covers. We select the neighboring document that is closest to the centroid of each cluster, and compile them to form the set of $k^*$ contrastive references: $\references_{\doc}$.

\input{tables/bright}

\input{tables/multi}
                                                    
\noindent\textbf{Contrastive Query Generation:}
For each contrastive reference document $\otherdoc{i}\in\references_{\doc}$, we use Claude-3.5-Sonnet~\citep{anthropicClaudeHaiku} to generate contrastive queries that are relevant to $\doc$ and not relevant to $\otherdoc{i}$. We combine the queries generated across all contrastive references to form $\queryset{\doc}$. We use the same prompt template for all datasets and do not engage in any prompt engineering (for the template and more details see Appendix~\ref{appendix:prompts}).

\noindent\textbf{Representation Sharpening:} During inference, we use Eq~\ref{eq:main} with a fixed $\alpha=1$, giving equal weight to the document and query components. 

\noindent\textbf{Baselines:}
We compare our method against the traditional retrieval pipeline. Additionally, we follow standard practice in query generation~\citep{ma-etal-2021-zero} and generate queries under a one-to-many paradigm i.e., from each individual document in the corpus (hereby called \textit{simple queries}). These queries form a stronger baseline: \textit{Doc2Query}, where the simple queries are concatenated to the end of the text of the document before embedding~\citep{nogueira2019document}. We benchmark these baselines against two variants of our method --- \textit{SimSharp}, where the queries $\queryset{d}$ used to shift the document in Eq~\ref{eq:main} are the simple queries, and \textit{ConSharp}, where $\queryset{d}$ is the set of contrastive queries. We use the standard BEIR metric of NDCG@10, with alternate metrics reported in Appendix~\ref{appendix:metrics}.

\noindent\textbf{Results:} 
Across all datasets, and retrievers (Table~\ref{table:beir}), \method{} consistently outperforms all baselines, with ConSharp achieving an average NDCG@10 improvement of 6.9\% over traditional inference. While both SimSharp and ConSharp provide significant gains, results show that using contrastive queries consistently boosts performance, validating the merit of our many-to-many query generation approach. This consistency underscores the generality of the method, and its ability to boost performance on a wide range of retrieval tasks, regardless of the underlying DR.

\subsection{Improving Previous Methods}
Recent work in zero-shot dense retrieval acts on the inference query $\query$. For example, HyDE~\citep{gao-etal-2023-precise} and Query2Doc~\citep{wang2023querydoc} prompt an LM to directly answer $\query$ and use this answer to obtain a refined query embedding $\queryembed{r}$. Follow-up work like LameR~\citep{shen2023large} advances answer generation, while ReDE-RF~\citep{jedidi2024zero} uses real documents instead of generated answers, but all maintain the fundamental approach of obtaining a richer $\queryembed{r}$ and computing document relevance with $\simscore(\queryembed{r}, \docembed)$ instead.  While these solutions operate on the query at inference time, our approach augments the document's representation. This makes these approaches compatible, as changing $g(\queryembed, \queryembedset{d})$ in Eq~\ref{eq:main} to  $g(\queryembed{r}, \queryembedset{d})$ combines \method{} with these methods. We use Qwen3-Embedding-0.6B as a dense retriever, and measure the performance of these methods when using traditional inference vs. ConSharp.

\noindent\textbf{Results:} When used in conjunction with previous methods, \method{} consistently outperforms traditional inference (Table~\ref{table:prior}), raising average NDCG@10 by 2.2\%. This shows that the framework synergizes well with prior work and can be seamlessly included to boost performance. 

%% file: tables/prior.tex
\begin{table*}[]
\centering
\begin{tabular}{@{}llrrrrrrr@{}}
\toprule
Method    & Inference   & FiQA           & NFCorpus       & SciFact        & Trec-COVID     & SciDocs        & Arguana        & Avg            \\ \midrule
HyDE      & Traditional & 37.88          & 34.47          & 73.86          & 86.41          & 21.90           & 55.63          & 51.69          \\
          & ConSharp    & \textbf{42.18} & \textbf{36.18} & \textbf{78.20}  & \textbf{86.51} & \textbf{22.61} & \textbf{57.02} & \textbf{53.78} \\\midrule
Query2Doc & Traditional & 41.04          & 34.97          & 74.70           & 86.62          & 22.37          & 65.70           & 54.23          \\
          & ConSharp    & \textbf{44.06} & \textbf{36.94} & \textbf{78.09} & \textbf{87.51} & \textbf{23.42} & \textbf{68.38} & \textbf{56.40}  \\\midrule
LameR     & Traditional & 39.29          & 38.55          & 73.85          & 83.14          & 21.96          & 50.63          & 51.24          \\
          & ConSharp    & \textbf{42.85} & \textbf{39.44} & \textbf{76.94} & \textbf{84.82} & \textbf{22.82} & \textbf{54.29} & \textbf{53.53} \\\midrule
ReDE-RF   & Traditional & 36.47          & 36.44          & 68.24          & \textbf{82.94} & 20.96          & 58.36          & 50.56          \\
          & ConSharp    & \textbf{38.94} & \textbf{37.93} & \textbf{72.22} & 82.63          & \textbf{22.95} & \textbf{60.93} & \textbf{52.60}  \\ \bottomrule
\end{tabular}
\caption{NDCG@10 when prior methods are deployed under traditional inference v.s. the \method{} framework with contrastive queries (ConSharp). ConSharp outperforms on all datasets, improving average NDCG@10 by 2.2\% and showing its consistent ability to boost the performance of prior methods.}
\label{table:prior}
\end{table*}

%% file: tables/bright.tex
\begin{table*}[]
\begin{tabular}{lrrrrrrrrrrr} \toprule
System       & \multicolumn{7}{c}{Stack Exchange}                                                                                                                                                                                                            & \multicolumn{2}{c}{Coding}                                     & \multicolumn{2}{c}{Theorem-Based}                  \\ 
             & Bio                                & Earth                     & Econ                               & Psy                      & Rob                                & Stack                              & Sus                                & Leet                               & Pony                      & AoPS           & TheoT                             \\ \midrule
Grit-LM      & 25.0                                 & 32.8                      & 19.0                                 & 19.9                     & 17.3                               & 11.6                               & 18.0                                 & 29.8                               & \textbf{22.0}               & 8.8            & 21.1                              \\
OpenAI       & 23.7                               & 26.3                      & 20.0                                 & 27.5                     & 12.9                               & 12.5                               & 20.3                               & 23.6                               & 2.5                       & 8.5            & 12.3                              \\
Voyage       & 23.6                               & 25.1                      & 19.8                               & 24.8                     & 11.2                               & 15.0                                 & 15.6                               & 30.6                               & 1.5                       & 7.4            & 11.1                              \\
Google       & 23.0                                 & \textbf{34.4}             & 19.5                               & 27.9                     & 16.0                                 & 17.9                               & 17.3                               & 29.6                               & 3.6                       & 9.3            & 14.3                              \\
Reason-IR-8B & 26.2                               & 31.4                      & 23.3                               & \textbf{30.0}              & 18.0                                 & 23.9                               & 20.5                               & 35.0                                 & 10.5                      & 14.7           & 27.2                              \\
+ ConSharp    & \textbf{27.5} & 31.5 & \textbf{23.5} & 29.9 & \textbf{18.5} & \textbf{25.7} & \textbf{21.7} & \textbf{36.1} & 10.8 & \textbf{14.9} & \textbf{28.3}  \\ \bottomrule
\end{tabular}
\caption{NDCG@10 on BRIGHT benchmark when \method{} is used on the ReasonIR-8B DR (+ ConSharp, other method results from ~\citet{shao2025reasonir}). ConSharp consistently boosts performance, leading to the ReasonIR-8B + ConSharp system establishing a new state-of-the-art on eight of eleven possible splits. }
\label{table:bright}
\end{table*}

%% file: tables/multi.tex
\begin{table}[]
\begin{tabular}{@{}llrrrr@{}}
\toprule
DR    & Variant               & Ko             & Th             & Hi             & Sw             \\ \midrule
MCont & Trad           & 29.31          & 40.73          & 14.58          & 32.25          \\
      & Doc2Q             & 21.07          & 18.18          & 8.54           & 33.59          \\
      & S.Sharp      & 39.89          & \textbf{52.91} & 23.75          & \textbf{47.04} \\
      & C.Sharp & \textbf{40.16} & 51.89          & \textbf{24.41} & 46.62          \\\midrule
E5-M  & Trad           & 47.05          & 61.82          & 30.71          & 61.34          \\
      & Doc2Q             & 44.87          & 56.27          & 30.65          & 59.72          \\
      & S.Sharp      & \textbf{54.82} & \textbf{67.83} & 40.61          & 65.64          \\
      & C.Sharp & 52.29          & 65.72          & \textbf{41.35} & \textbf{66.76} \\ \bottomrule
\end{tabular}
\caption{NDCG@10 on MIRACL datasets. Representation sharpening consistently improves performance, showing that the method can boost dense embedders on tasks that span across languages.}
\label{table:multi}
\end{table}

%% file: sections/3_2_bright.tex
\subsection{State-of-the-art on retrieval for reasoning}

BEIR consists of information-seeking queries where DRs perform well. However, complex queries may require reasoning that goes beyond surface form matching~\citep{su2025bright}. The BRIGHT benchmark consists of such queries, with domains such as arithmetic and coding. We use ReasonIR-8B~\citep{shao2025reasonir}, a DR which achieves state-of-the-art performance on BRIGHT, and measure the performance increase when using \method{} by contrastive queries (ConSharp). We set $\alpha=0.2$, based on a hyperparameter search in the range $\alpha\in[0.05, 1.5]$ on one of BRIGHT's subsets (TheoremQA Questions, arbitrarily chosen) and omit this subset from the results.

\noindent\textbf{Results:} Representation sharpening consistently boosts ReasonIR on BRIGHT (Table~\ref{table:bright}). When combined, the new system achieves state-of-the-art performance on eight of eleven measured subsets. This shows that the framework can be helpful even in complicated domains like reasoning or coding. 

%% file: sections/3_3_miracl.tex
\subsection{Boosting multilingual retrievers}
Representation sharpening leads to improvements on a range of English-language datasets. However, given that LMs are known to perform better on English than on other languages~\citep{asai-etal-2024-buffet}, it is unclear whether the method can perform well in the multilingual case. To test this, we select four diverse low-resource languages from the MIRACL benchmark (Korean, Hindi, Swahili, Thai) and use the multilingual DRs of MContriever~\citep{izacard2022unsupervised} and E5-Multilingual-Large-Instruct~\citep{wang2024multilingual}. Since Claude-3.5-Sonnet has shown impressive multilingual capabilities~\citep{enis2024llm}, we do not change the generating model. We use the same English prompt template from previous experiments, as the LM proves capable of generating contrastive queries in the appropriate language for each document.

\noindent\textbf{Results}: 
Regardless of the language or DR, \method{} offers consistent and considerable performance boosts (Table~\ref{table:multi}), with ConSharp improving average NDCG@10 by 8.9\%. Surprisingly, while sharpening with simple queries (SimpSharp) delivers fewer gains on the BEIR datasets (Table~\ref{table:beir}), it performs competitively in the multilingual setting, suggesting that LMs like Claude-3.5-Sonnet lack the power to create nuanced contrastive queries in non-English languages. 

\input{tables/approx}

%% file: tables/approx.tex
\begin{table*}[]
\centering
\begin{tabular}{@{}lrrrrrrr@{}}
\toprule
Variant     & FiQA           & NFCorpus      & SciFact        & Trec-COVID     & SciDocs        & Arguana       & Avg            \\ \midrule
Traditional & 34.28          & 30.41         & 64.61          & 58.49          & 16.48          & 66.60          & 45.15          \\
DocExp     & 34.81          & 30.85         & 64.65          & \textbf{71.32} & 20.14          & 64.95         & 47.78          \\
IndexSharp   & \textbf{40.74} & \underline{33.80} & \underline{69.73} & 62.73          & \underline{22.45} & \underline{69.60} & \underline{49.84} \\ \midrule
ConSharp     & \underline{40.62}          & \textbf{34.30}          & \textbf{71.88}          & \underline{62.90}           & \textbf{22.55}          & \textbf{69.97}         & \textbf{50.37}          \\ \bottomrule
\end{tabular}
\caption{NDCG@10 of ConSharp when compared (\textbf{best}, \underline{second-best}) to approaches that offload all cost to the indexing operation. Indexing time sharpening (IndexSharp) consistently outperforms other index-time baselines, and captures the majority of ConSharp's performance increases while incurring no additional inference-time costs.}
\label{table:approx}
\end{table*}

%% file: sections/5_approximation.tex
\section{Index-Time Sharpening}
\label{sec:approx}
The \method{} framework provides consistent performance gains on a wide variety of domains and languages, however it comes at a cost. Each document is paired with a set of contrastive queries $\queryset{d}$, the embeddings of which are stored in the index as meta-data. This leads to the index growing by a factor of $\frac{1}{|\corpus|}\sum_{\doc\in\corpus}|\queryset{\doc}|$. Seeking to eliminate any inference time cost to the algorithm, we identify strategies by which we may use  contrastive queries to augment a document's representation during the indexing phase alone. 

The first strategy is document expansion (DocExp)~\citep{nogueira2019document}; for every document $\doc$, we generate contrastive queries $\queryset{d} = \{\query{\doc, 1}, \query{\doc, 2}, \ldots\}$ and expand the document by concatenating the text of the queries: $\doc{\text{expand}} = d\circ\query{\doc, 1}\circ\query{\doc, 2}\ldots$. The index then stores the embedding of this expanded document $\docembed{\text{expand}}$ and the relevance score is given by $\simscore(\queryembed, \docembed{\text{expand}})$. 

We further introduce a novel method of more directly editing the index, \textit{IndexSharp}. We take inspiration from the parameter-free adapter (PEFA)~\citep{chang2024pefa}, which selects training set queries and uses them to augment the document's representation in the index, and extend the method to the zero-shot setting. We set $g$ in Eq~\ref{eq:main} to $\frac{1}{|\queryset{d}|}\sum_{\queryembed{d}\in\queryembedset{d}}\queryembed{d}$, removing dependence on inference queries and isolating cost to indexing. We implement both of these alternatives, using Qwen3-Embedding-0.6B as the underlying DR.

\noindent\textbf{Results:} 
All approaches consistently outperform the traditional retrieval pipeline (Table~\ref{table:approx}), with IndexSharp improving average NDCG@10 by 4.7\%. While ConSharp method (inference-time) is typically superior, it is encouraging to see that the performance gap is not wide, suggesting that the bulk of our performance improvements can be preserved with no inference time cost.

%% file: sections/4_ablations.tex
\section{Ablations and Analysis}
\label{sec:ablations}
We conduct ablations and analysis with the goal of better understanding which components and design decisions most contribute to the method's success. 

\input{figures/lm} 

\noindent\textbf{Importance of Document Quality:} For each test query, we retrieve the top 100 documents and compute each document's effective \textit{sharpening boost}: $\simscore{}(\queryembed, \docembed^*)-\simscore(\queryembed, \docembed)$. A high sharpening boost implies that using \method{} leads to the relevance score increasing by a significant amount. We then follow prior work~\citep{Dathathri2020Plug} and compute the perplexity as a measure of document fluency, and observe the Pearson correlation ~\citep{pearson1895vii} between these two values. Across all datasets, we see a negative correlation (Table~\ref{table:quality}), suggesting that the more fluent a document is, the higher its sharpening boost. The result stresses the importance of crafting documents that are of high quality, as this may effect the boost they receive.

\input{tables/quality}

\input{figures/alpha}

\input{figures/nqs}

\noindent\textbf{Language Model:} With the SciFact dataset and the Qwen3-Embedding-0.6B DR, we vary the LM that generates the contrastive queries, using Claude-3-Haiku and open-weight models from the Qwen3~\citep{zhang2025qwen3} family. The results (Figure~\ref{fig:lm}) show that performance scales with the size of the model. However, even Qwen3-8B gives a significant performance boost, showing that \method{} is not reliant on the largest of models, and can be useful even when powered by smaller, more accessible open weight models. Encouragingly, the performance gap between the Qwen3-32B model and Claude-3-5-Sonnet is only 1.35\%, suggesting that a sufficiently powerful open-source model can enable similar performance gains as the most capable frontier models.

\noindent\textbf{Hyperparameters:} Our framework consists of hyperparameter $\alpha$, that controls the weight between the document and the query component of the sharpened representation, and an implicit $n$, i.e., the number of contrastive queries used (per document) during inference. We fix the DR to the E5-Mistral model and vary these two. We see (Figure~\ref{fig:alpha}) that increasing $\alpha$ from $0$ to $1.5$ leads to significant improvements in performance, after which performance declines. The peak performance for each dataset is never achieved by $\alpha=1$, which is the value we use in our experiments. This shows that, should a validation set be available, \method{} could be further improved by tuning the value of $\alpha$. Increasing the number of queries used (Figure~\ref{fig:nqs}) is always beneficial. With $n>10$, performance begins to plateau, suggesting that 10 queries are enough for the system to reach high levels of performance.

%% file: figures/lm.tex
\begin{figure}[th]
    \centering
    \includegraphics[width=\linewidth]{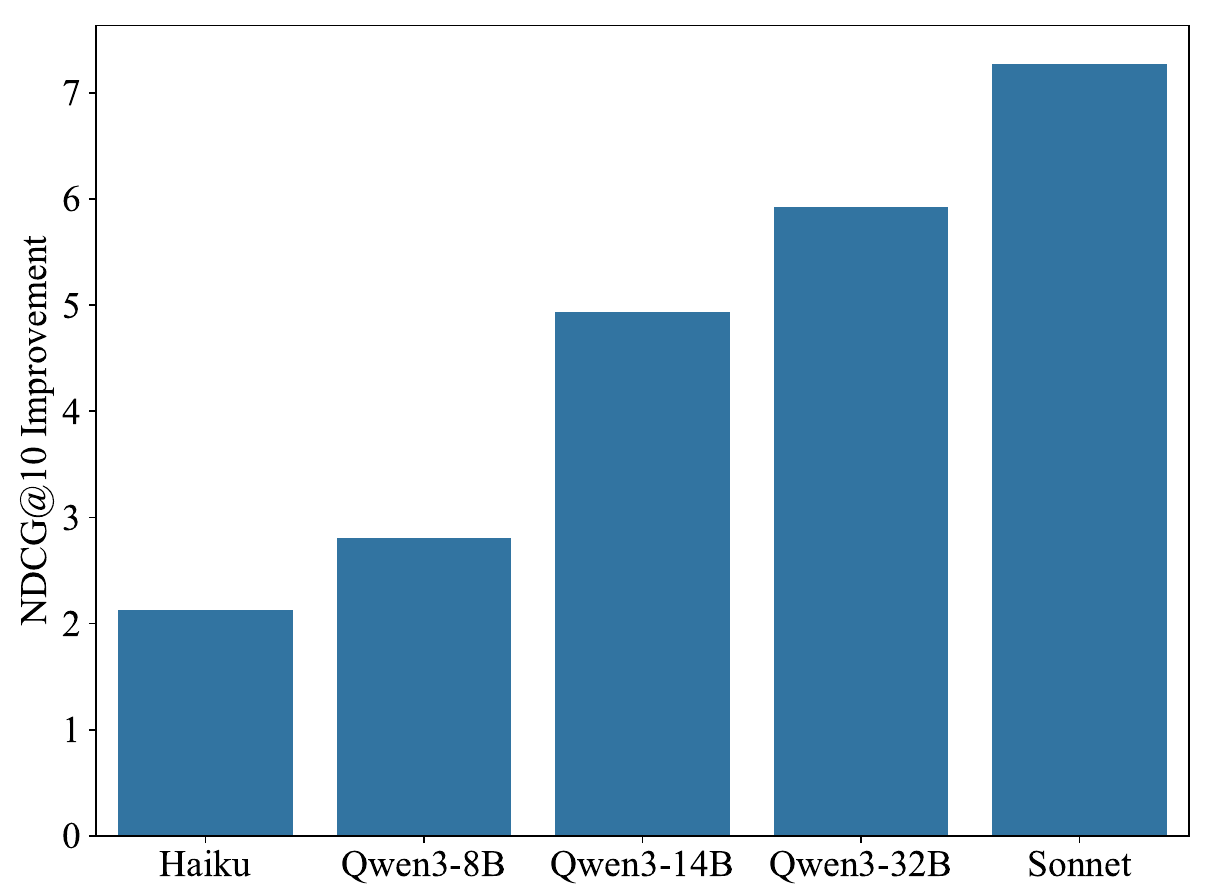}
    \caption{Performance improvement on SciFact when varying the underlying LM. Performance scales with model size, however Qwen3-8B still delivers a noteably performance boost, showing that even open-weight models on the 8B parameter scale can be used to generate effective contrastive queries.}
    \label{fig:lm}
\end{figure}

%% file: tables/quality.tex
\begin{table*}[]
\centering
\begin{tabular}{@{}lrrrrrrr@{}}
\toprule
DR         & FiQA   & NFCorpus & SciFact & Trec-COVID & SciDocs & Arguana & Avg   \\ \midrule
Contriever & -2.36 & -5.81    & -13.71 & -6.77     & -2.19  & -6.19  & -6.18 \\
Qwen3      & -1.88  & -13.85   & -8.45   & -3.81      & -4.94   & -11.75  & -7.45 \\
E5-Mistral & -4.29  & -3.52    & -8.72   & -9.35      & -2.86   & -9.36   & -6.35 \\ \bottomrule
\end{tabular}
\caption{Pearson correlation between the perplexity and sharpening boost received by documents. More fluent documents receive greater boosts, highlighting the important of creating high quality and fluent documents.}
\label{table:quality}
\end{table*}

%% file: figures/alpha.tex
\begin{figure}[t]
    \centering
    \includegraphics[width=\linewidth]{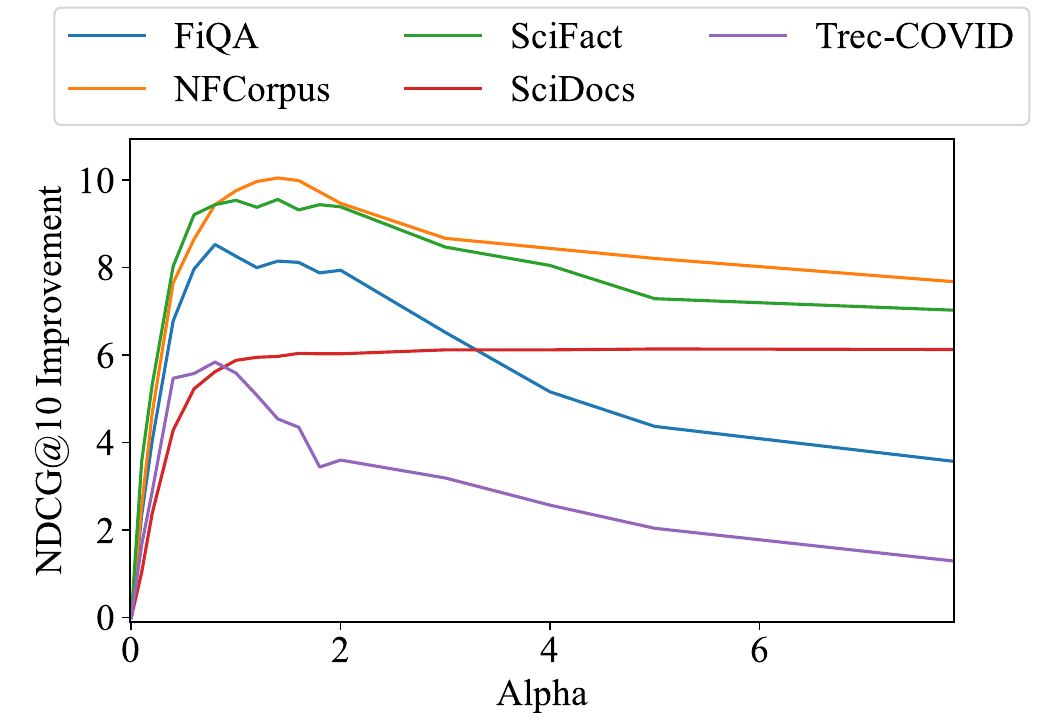}
    \caption{Performance when varying $\alpha$ on E5-Mistral. Increasing $\alpha$ from 0 to 1 always leads to improvements and performance declines after peaking in the $\alpha\in[1, 1.5]$ range. The best value of $\alpha$ is never the value used in our experiments ($\alpha=1$), suggesting that hyperparameter tuning can provide further gains.}
    \label{fig:alpha}
\end{figure}

%% file: figures/nqs.tex
\begin{figure}[th]
    \centering
    \includegraphics[width=\linewidth]{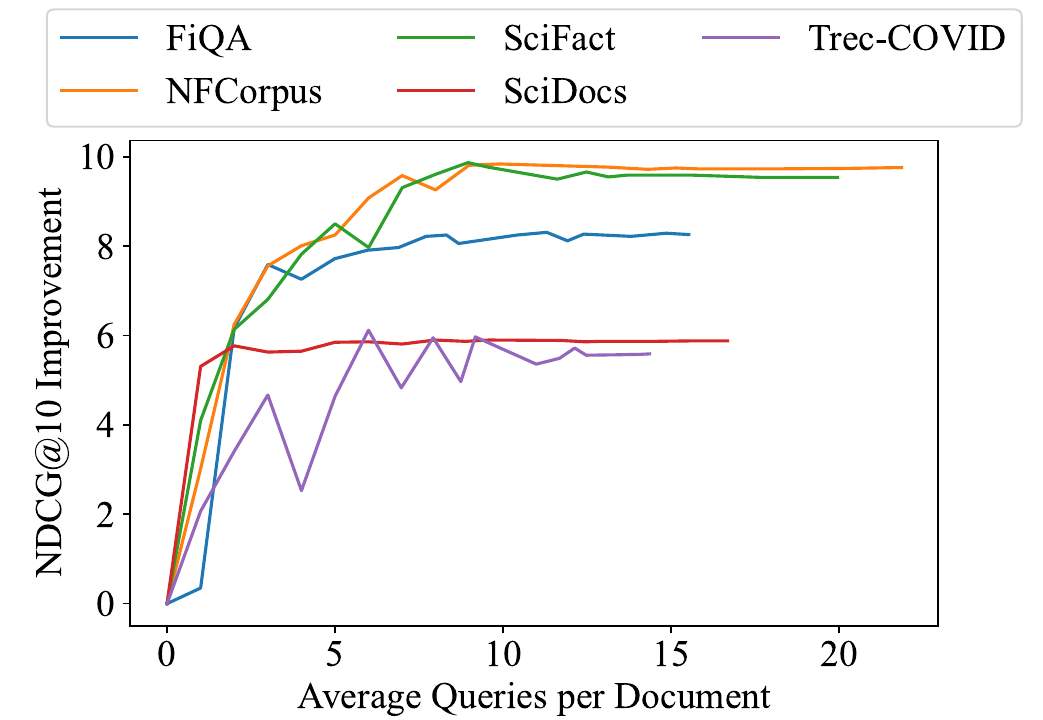}
    \caption{Performance when varying the average number of queries used per document on E5-Mistral. Using more queries increases performance, with the majority of the improvement occurs with the first $10$ queries.}
    \label{fig:nqs}
\end{figure}

%% file: sections/conclusion.tex
\section{Conclusion}
\label{sec:conclusion}
We tackle the challenging setting of zero-shot dense retrieval, and propose a framework for \textbf{\method{}} that boosts the performance of a DR without retraining. To operationalize our framework, we introduce a \textbf{many-to-many} paradigm for query generation, deploying it to create \textbf{contrastive queries} that accentuate the aspects that distinguish a document from other, similar documents in the corpus. Through extensive experimentation, we show that \method{} outperforms traditional retrieval inference on a variety of tasks and languages, boosts prior methods and sets a new state-of-the-art on the BRIGHT benchmark. Finally, we devise an indexing time algorithm that achieves considerable performance gains, showing that \method{} can boost performance with \textbf{no additional inference-time cost}.

%% file: sections/limitations.tex
\section{Limitations}
Much like prior work in zero-shot dense retrieval~\citep{gao-etal-2023-precise, wang2023querydoc}, the \method{} framework relies on Language Models, using them for query generation. This imposes a natural limitation on the range of domains where the method can be expected to deliver gains, based on the capability of the LM used. In domains where the LM suffers from a lack of comprehension (an unknown language or niche and technical context), we would not expect our method to perform well. Additionally, as is common with methods in training-free adaptation~\cite{gao-etal-2023-precise, chang2024pefa}, while we are able to improve performance, the extent of the improvement does not match that of training a superior DR. This can be seen in Table~\ref{table:beir}, where, though ConSharp outperforms the traditional baseline on the Contriever model, it is generally not sufficient to make the Contriever model better than the Qwen3-Embedding-0.6 model. This suggests that while the method can offer significant performance gains, it cannot replace  a superior underlying retrieval system. Finally, while we limit our scope to training-free approaches, contrastive query generation naturally produces contrastive triplets of the form $(\queryset{\doc}, \doc, \otherdoc)$ and future work may explore the merits of using such queries for contrastive learning~\citep{izacard2022unsupervised}. 

\section{Ethical Considerations}
This work leverages LMs for synthetic data generation, raising several important ethical considerations. Recent studies have found LMs to contain a variety of implicit biases on axes spanning gender, race, political affiliation, etc.~\citep{resnik2025large}. Practitioners must consider their specific choice of generating model, and investigate how potential biases could affect the queries generated. Additionally, the most powerful LMs are often closed-source, API-access models~\citep{anthropicClaudeHaiku, brown2020language}, as opposed to open-weight models that can be run on internal hardware. To use the \method{} framework with these models, one must send data through API calls, a procedure which could expose confidential data to external actors. To ensure an equitable use of the method, practitioners must consider the privacy protocols of the inference API service under consideration, and ask whether the corpus used for query generation contains sensitive information. 

%% file: sections/A_appendix.tex
\section{Alternate Metrics}
\label{appendix:metrics}

Aside from the traditional metric of NDCG@10, we also provide measurements of Recall@50 (Table~\ref{fig:alpha}) and MAP@50 (Table~\ref{table:beirmap}), to show that the consistent outperformance of \method{} is robust across metrics. 

\input{tables/beir_recall}

\input{tables/beir_map}

\noindent Negative Contrastive Query Examples: 
\begin{verbatim}
Document 1: Statin use after diagnosis 
of breast cancer and survival: a 
population-based cohort study...

Document 2: Dietary intakes of mushrooms
and green tea combine to reduce the risk
of breast cancer in Chinese women. ....

Contrastive Negative Queries:
1. mushrooms breast cancer prevention
2. green tea breast cancer
3. Chinese women diet cancer
\end{verbatim}

\section{Contrastive Query Generation Prompts}
\label{appendix:prompts}
Given a document text $d$, we are provided with a contrastive reference document $\otherdoc$ and tasked with generating contrastive queries that are relevant to $\doc$ but not relevant to $\otherdoc$. We achieve this using the following prompt:

\begin{verbatim}
Here are some examples of queries to
understand the style:
[EXAMPLE 1]
[EXAMPLE 2]
[EXAMPLE 3]
[EXAMPLE 4]
[EXAMPLE 5]

Given the following two documents, 
create a query that is weakly 
related to both documents such that 
document 1 is directly relevant to 
the query, but document 2 is not. 
Your plan should highlight the key 
difference between the documents 
that you will use. 

Document 1: [DOCUMENT d]
Document 2: [Document d']
Return as many answers as you can, 
but make sure that each answer is 
unique and distinct. Do not repeat 
yourself across answers, and focus on 
quality over quantity. 
Format your answer in the following 
output structure:

<PLAN>Explanation on how you will design 
the query and why the first document is
relevant to it but the second is not. 
Also explain how you will ensure the 
style is similar to the style of the 
queries provided above 
(name the language you will use)</PLAN>

<QUERY>text of the query in the same 
language as the examples and 
document above</QUERY>    
\end{verbatim}

We use the \textbf{same} 5 example queries for all ($\doc, \otherdoc$) pairs of a given dataset, ensuring our method does not require access to more than 5 queries from the domain of interest. 

For the Arguana dataset alone, we ask for a `counter-argument passage' instead of a `query', keeping all other parts of the prompt the same.

We then parse the output text using regular expressions to extract the generated queries. 

\subsection{Example Outputs: }

\noindent FiQA Dataset
\begin{verbatim}
Document 1: Just have the associate 
sign the back and then deposit it.  
It's called a third party cheque and 
is perfectly legal.  I wouldn't be 
surprised if it has a longer hold 
period and, as always, you don't get
the money if the cheque doesn't clear.
Now, you may have problems if it's 
a large amount or you're not very 
well known at the bank.  In that 
case you can have the associate 
go to the bank and endorse it in 
front of the teller with some ID.  
You don't even technically have to 
be there.  Anybody can deposit money 
to your account if they have the 
account number. He could also just 
deposit it in his account and write 
a cheque to the business.

Document 2: "Lets say you owed me
$123.00 an wanted to mail me a 
check. I would then take the check
from my mailbox an either take 
it to my bank, or scan it and 
deposit it via their electronic 
interface. Prior to you mailing 
it you would have no idea which 
bank I would use, or what my 
account number is. In fact I 
could have multiple bank accounts, 
so I could decide which one to 
deposit it into depending on what 
I wanted to do with the money, 
or which bank paid the most 
interest, or by coin flip. Now 
once the check is deposited my 
bank would then ""stamp"" the 
check with their name, their 
routing number, the date, an my 
account number. Eventually an 
image of the canceled check would 
then end up back at your bank. 
Which they would either send to you, 
or make available to you via 
their banking website. You don't 
mail it to my bank. You mail it 
to my home, or my business, or 
wherever I tell you to mail it. 
Some business give you the address 
of another location, where either 
a 3rd party processes all their 
checks, or a central location  
where all the money for multiple 
branches are processed. If you 
do owe a company they will generally 
ask that in the memo section in 
the lower left corner that you 
include your customer number. 
This is to make sure that if they 
have multiple Juans the money 
is accounted correctly. In all 
my dealings will paying bills 
and mailing checks I have never 
been asked to send a check 
directly to the bank. If they 
want you to do exactly as you 
describe, they should provide 
you with a form or other instructions."

Contrastive Queries: 
1. How do you deposit a third-party 
check at a bank?
2. Is endorsing a check in front of a 
teller necessary for deposit?
3. Can you deposit money into someone 
else's account with just their 
account number?
\end{verbatim}

\noindent NFCorpus Dataset
\begin{verbatim}
Document 1: Statin use after diagnosis 
of breast cancer and survival: a 
population-based cohort study...

Document 2: Dietary intakes of mushrooms
and green tea combine to reduce the risk
of breast cancer in Chinese women. ....

Contrastive Queries:
1. statins breast cancer
2. simvastatin
3. Cohort studies breast cancer
\end{verbatim}

\section{Datasets}
\label{appendix:datasets}

We use the following datasets in our experiments. All datasets / splits used are available under permissive licenses that are in line with our use case.

\noindent\textbf{BEIR Benchmark:}
From the BEIR benchmark~\citep{thakur2021beir}, we select all datasets that have under 200K documents in the corpus:
\begin{itemize}
    \item FiQA~\citep{maia201818}: A financial question answering dataset with 57K documents and 648 test queries
    \item NFCorpus~\citep{boteva2016full}: A dataset for biomedical IR with 3.6K	documents and 323 test queries
    \item SciFACT~\citep{wadden2022scifact}: A dataset for scientific fact verification with 5K documents and 300 test queries
    \item Trec-COVID~\citep{roberts2021searching}: A dataset for scientific retrieval of COVID related documents with 171K documents and 50 test queries
    \item SciDocs~\citep{cohan-etal-2020-specter}: A dataset of with 25K documents and 1000 test queries
    \item Arguana~\citep{wachsmuth2018retrieval}: A dataset for counterargument mining with 8.67K documents and 1406 test queries. 
\end{itemize}

\noindent\textbf{MIRACL Benchmark:}

We select four of the lowest resource languages from the MIRACL~\citep{zhang-etal-2023-miracl} benchmark, ensuring to maintain diversity across regions and language families:

\begin{itemize}
    \item Hindi: 148,107 documents 350 test queries 
    \item Korean: 437,373 documents 213 test queries
    \item Swahili: 47,793 documents 482 test queries
    \item Thai: 128,179 documents 733 test queries
\end{itemize}

\noindent\textbf{BRIGHT Benchmark:}

We use all splits of the BRIGHT~\citep{su2025bright} benchmark. 

\section{Hardware}
\label{appendix:hardware}
Experiments were run on two systems:

\begin{enumerate}
    \item System 1: Most experiments. 32 CPUs, 244 GiB memory, Processor: Intel Xeon E5-2686 v4 (Broadwell), Clock Speed: 2.3 GHz and 4 NVIDIA-TeslaV100 GPUs each with 16GB GPU memory.
    \item System 2: For only the most compute intensive jobs (Qwen3 LM ablation). 24 CPUs, 1152 GiB memory, Processor: Intel Xeon Platinum 8275L, Clock Speed: 3 GHz and 8 NVIDIA-A100 GPUs, each with 40GB GPU memory. 
\end{enumerate}

%% file: tables/beir_recall.tex
\begin{table*}[th]
\centering
\begin{tabular}{@{}llrrrrrr@{}}\toprule
DR         & Variant         & FiQA           & NFCorpus       & SciFact        & Trec-COVID    & SciDocs        & Arguana        \\\midrule
Contriever & Trad            & 46.40           & 23.34          & 86.70           & 1.59          & 29.15          & 94.24          \\
           & Doc2Query       & 40.95          & 23.52          & 86.70           & 2.31          & 29.36          & 95.31          \\
           & SimSharp      & \textbf{51.09} & 24.52          & 89.93          & 1.81          & 31.62          & 96.23          \\
           & ConSharp & \textbf{51.83} & \textbf{24.77} & \textbf{90.17} & \textbf{1.97} & \textbf{32.35} & \textbf{96.51} \\\midrule
Qwen3      & Trad            & 60.67          & 24.48          & 88.89          & 6.04          & 36.07          & 98.51          \\
           & Doc2Query       & 59.64          & 24.34          & 88.56          & 6.18          & 38.30           & 98.22          \\
           & SimSharp      & 63.04          & \textbf{25.68} & 90.67          & \textbf{6.68} & 40.71          & 98.58          \\
           & ConSharp & \textbf{64.32} & \textbf{26.55} & \textbf{90.89} & 6.62          & \textbf{41.14} & \textbf{98.86} \\\midrule
E5-Mistral & Trad            & 59.58          & 23.49          & 89.00             & 3.36          & 6.50            & 94.24          \\
           & Doc2Query       & 58.86          & 20.18          & 88.67          & \textbf{14.01}         & 3.39           & 93.74          \\
           & SimSharp      & 64.89          & 25.79          & 90.83          & 3.66          & 8.99           & 95.80           \\
           & ConSharp & \textbf{66.95} & \textbf{26.64} & \textbf{91.50}  & 3.75 & \textbf{9.05}  & \textbf{95.92}\\\bottomrule
\end{tabular}
\caption{Recall@50 on BEIR benchmark. Across datasets and underlying dense retrievers, \method{} outperforms traditional inference. }
\label{table:beirrecall}
\end{table*}

%% file: tables/beir_map.tex
\begin{table*}[th]
\centering
\begin{tabular}{@{}llrrrrrr@{}}\toprule
DR         & Variant         & FiQA           & NFCorpus       & SciFact        & Trec-COVID    & SciDocs        & Arguana        \\\midrule
Contriever & Trad            & 19.92          & 14.18          & 53.29          & 0.93          & 9.71           & 42.10           \\
           & Doc2Query       & 16.19          & 14.46          & 53.49          & \textbf{1.64}          & 9.73           & 42.55          \\
           & SimSharp      & \textbf{23.98} & 15.01          & 61.25          & 1.27          & 11.53          & 45.95          \\
           & ConSharp & 23.91          & \textbf{15.91} & \textbf{64.06} & 1.35 & \textbf{12.10}  & \textbf{49.31} \\\midrule
Qwen3      & Trad            & 28.53          & 13.45          & 60.48          & 6.65          & 11.37          & 58.98          \\
           & Doc2Query       & 29.53          & 13.69          & 60.35          & 7.30           & 14.01          & 57.45          \\
           & SimSharp      & 32.34          & 15.03          & 65.8           & \textbf{7.30}  & 15.04          & 58.91          \\
           & ConSharp & \textbf{34.70}  & \textbf{15.52} & \textbf{66.05} & 7.13          & \textbf{15.67} & \textbf{62.89} \\\midrule
E5-Mistral & Trad            & 31.11          & 10.95          & 60.58          & 2.64          & 1.64           & 43.08          \\
           & Doc2Query       & 29.32          & 8.51           & 60.71          & \textbf{3.97}          & 2.58           & 40.61          \\
           & SimSharp      & 36.70           & 15.05          & 67.92          & 2.93          & 4.17           & 48.84          \\
           & ConSharp & \textbf{38.84} & \textbf{16.22} & \textbf{70.52} & 3.02 & \textbf{5.22}  & \textbf{49.41}\\\bottomrule
\end{tabular}
\caption{MAP@50 on BEIR benchmark. Across datasets and underlying dense retrievers, \method{} outperforms traditional inference. }
\label{table:beirmap}
\end{table*}